\documentclass[11pt]{article} 
\usepackage{amsmath}
\usepackage{mathrsfs}
\usepackage{amsfonts}
\usepackage{graphicx}
\usepackage{amssymb}
\usepackage[utf8]{inputenc} 
\usepackage[T1]{fontenc}
\usepackage{color}
\newcommand{\be}{ \begin{equation}}
\newcommand{\ee}{\end{equation}} 
\begin{document} 
\def\theequation{\arabic{section}.\arabic{equation}} 
\begin{titlepage} 
\title{Cosmic analogues of classic variational problems} 
\author{Valerio Faraoni\\ \\ 
{\small \it Department of Physics \& Astronomy, Bishop's University}\\ 
{\small Sherbrooke, Qu\'ebec, Canada J1M~1Z7} }
\date{} \maketitle 
\thispagestyle{empty} 
\vspace*{1truecm} 

\begin{abstract} 

Several classic one-dimensional problems of variational calculus 
originating in non-relativistic particle mechanics have solutions that are 
analogues of spatially homogeneous and isotropic universes. They are ruled 
by an equation which is formally a Friedmann equation for a suitable 
cosmic fluid. These problems are revisited and their cosmic analogues are 
pointed out. Some correspond to the main solutions of cosmology, while 
others are analogous to exotic cosmologies with phantom fluids and finite  
future singularities.

\end{abstract} 
\vspace*{1truecm} 
%\begin{center} PACS: 04.20.Cv, 04.30.-w
% fundamental problems ..., grav. waves: theory
%{\bf Keywords:} scalar-tensor theories 
%\end{center} 
\end{titlepage}

\def\theequation{\arabic{section}.\arabic{equation}}

%%%%%%%%%%%%%%%%%%%%%%%%%%%%%%%%%%%%%%%%%%%%%%%%%%%%%%%%%%%%%%%%%%%%%%%%%%%%%%%%%%%%

\section{Introduction}
\setcounter{equation}{0}
\label{sec:1}

Several classic one-dimensional problems of mechanics solved with 
variational calculus 
have analogues in spatially homogeneous and isotropic (or 
Friedmann-Lema\^itre-Robertson-Walker, hereafter ``FLRW'') cosmology. The 
analytical solutions of these problems often correspond to particularly 
important solutions of FLRW cosmology. One can build in the laboratory 
many physical systems that are analogous to curved spacetimes describing 
black holes or universes, using which one can study curved space quantum 
effects such as Hawking radiation, particle creation, or superradiance. 
These systems include Bose-Einstein condensates and other condensed matter 
systems \cite{BEC1, BEC2, BEC3, BEC4, BEC5, BEC6, BEC7, BEC9, CM1, 
CM2, CM3, CM4, CM5, CM6}, fluid-dynamical systems \cite{fluid1,fluid2, 
fluid3, fluid4, fluid5, fluid6, fluid7, fluid8, fluid9, fluid10, fluid11, 
fluid12, fluid13}, and optical systems \cite{optical1, optical2, optical3, 
optical4, optical5} and they have originated the field of research known 
as analogue gravity ({\em e.g.}, \cite{analoguegeneral1, analoguegeneral2, 
analoguegeneral3, analoguegeneral4, analoguegeneral5}), part of which 
focuses on analogues of FLRW cosmology in Bose-Einstein condensates 
\cite{acosmo1, acosmo2, acosmo3, acosmo4, analoguegeneral2, BEC2, BEC3, 
BEC4, BEC5, BEC6, BEC7, CM2}. Cosmic analogue systems can also be created 
with soap bubbles \cite{CriadoAlamo,bubble2} and capillary fluid flow 
\cite{capillary}. Less known analogues for cosmology, which involve 
natural systems outside of the laboratory, include glacial valley profiles 
\cite{Gibbons0, FACETS}, equilibrium beach profiles \cite{beach}, the 
freezing of bodies of water \cite{freezing}, and the Omori-Utsu law for 
the aftershocks following a main earthquake shock \cite{myOmori}.

Let us go over the basic concepts of FLRW cosmology to fix the notations 
and the terminology. The Einstein equations read\footnote{We use units in 
which the speed of 
light is unity and we follow the notations of Ref.~\cite{Wald}.} 
\cite{Wald, Carroll, Liddle}
\be
{\cal R}_{ab}-\frac{1}{2} \, g_{ab} {\cal R} +\Lambda g_{ab}=8\pi G 
T_{ab} \,,\label{einsteineqs}
\ee
where $G$ is Newton's constant, $g_{ab}$ is the spacetime metric, ${\cal 
R}_{ab}$ is the Ricci tensor, ${\cal R}={{\cal R}^a}_a$ is the Ricci 
scalar, 
$\Lambda$ is the cosmological constant, and $T_{ab}$ is the stress-energy 
tensor of matter. 

The FLRW line element is\footnote{The fact that the scale factor depends 
only on time ({\em i.e.}, the high degree of symmetry of FLRW spaces) 
requires analogue problems to be one-dimensional.}  
\be
ds^2 = -dt^2 +a^2(t) \left[ \frac{dr^2}{1-Kr^2} +r^2 \left( d\theta^2 + 
\sin^2 \theta \, d\varphi^2 \right)\right]  \label{eq:10}
\ee
in comoving polar coordinates $\left(t, r, \theta, \varphi  \right)$, 
where the curvature index $K$ can be normalized to $K= 1, 0, -1$ (although 
this is not necessary), 
corresponding to a universe with closed 
three-dimensional  spatial sections, Euclidean spatial 
sections, or hyperbolic 
3-sections, respectively \cite{Carroll,Wald,Peebles,Liddle,KT}.

The matter content of the universe causing the spacetime curvature is 
usually modelled by a perfect fluid with stress-energy tensor
\be
T_{ab}=\left( P+\rho\right) u_a u_b +P g_{ab} \,,
\ee
where $u^a $ is the four-velocity of the fluid and of comoving observers, 
while the energy 
density $\rho(t)$ and isotropic pressure $P(t)$ are related by some 
equation of state. Usually the latter has the barotropic form $P=P(\rho)$ 
 and often (but not necessarily) $P=w\rho$ with $w=$~const. Formally, the 
cosmological constant term can be treated as a 
special case of a perfect fluid with effective stress-energy tensor 
$ T_{ab}^{(\Lambda)} =- \Lambda g_{ab}/(8\pi G) $ with effective equation 
of state  $ 
P_{\Lambda}= - \rho_{\Lambda}= - \frac{\Lambda}{8\pi G}$.

The scale factor $a(t)$ of the FLRW metric~(\ref{eq:10}), the energy 
density $ \rho(t)$ 
and the pressure $P(t)$ obey the Einstein-Friedmann equations
\begin{eqnarray}
&&H^2 \equiv \left( \frac{\dot{a}}{a}\right)^2 =\frac{8\pi G}{3} \, \rho 
-\frac{K}{a^2} \,, \label{eq:11}\\
&&\nonumber\\
&&\frac{\ddot{a}}{a}= -\, \frac{4\pi G}{3} \left( \rho +3P \right) \,, 
\label{eq:12} \\
&&\nonumber\\
&& \dot{\rho}+3H\left(P+\rho \right)=0 \,,\label{eq:13}
\end{eqnarray}
where an overdot denotes differentiation with respect to the comoving time 
$t$ and $H(t)\equiv \dot{a}/a$ is the Hubble function 
\cite{Carroll,Wald,Peebles,Liddle,KT}. Only two equations in this set are 
independent:  given any two of them, the third one can be derived from the 
others. Without loss of generality, we choose the Friedmann 
equation~(\ref{eq:11}) and the energy conservation equation~(\ref{eq:13}) 
as independent, while the acceleration equation~(\ref{eq:12}) follows from 
them.

If the cosmic fluid satisfies the barotropic equation of state $ P=w\rho $ 
with $w=$~const., the covariant conservation equation~(\ref{eq:13})  
integrates to
\be
\rho(a) = \frac{ \rho_0}{ a^{3(w+1)} }  \label{eq:16} \,.
\ee
Further, if the universe is spatially flat ({\em i.e.}, $K=0$), the 
corresponding scale factor is
\be
a(t)=\frac{a_0}{ \left| t-t_0\right|^{3|w+1| } } \,.
\ee

Solution methods, phase space, and analytic solutions of the 
Einstein-Friedmann equations are reviewed in \cite{oldAmJP, AmJP1, 
SonegoTalamini}), while \cite{Chen0, Gibbons0, roulettes} report new 
results. In particular, a mathematical property of the Friedmann 
equation~(\ref{eq:11}) relevant here and proved in~\cite{roulettes} is 
that the graphs of all solutions of this equation are roulettes. A 
roulette is the locus of a point that lies on, or inside, a curve that 
rolls without slipping along another given curve. Special cases include 
the elliptical cycloid, which is the curve described by a point on an 
ellipse as the latter rolls on the $x$-axis. When the ellipse reduces to a 
circle, one reproduces the ordinary cycloid (the trajectory of a point on 
the rim of  a bicycle wheel as the bicycle advances at constant speed on 
horizontal ground) and is probably the most well known roulette.
 
It may seem that a complete analogy between the Einstein-Friedmann 
equations and the physical systems that we will consider is not complete 
because in the latter the dynamics is described by a single differential 
equation (analogous to the Friedmann equation), while the universe is 
described by two equations of the set~(\ref{eq:11})-(\ref{eq:13}). This is 
not the case because the information contained in the second equation 
(say, the covariant conservation equation (\ref{eq:13})) has already been 
inserted into the first one (the Friedmann equation (\ref{eq:11}) by using 
the scaling~(\ref{eq:16}) of the perfect fluid energy density and of the 
curvature term with the scale factor $a$ on the right hand side of 
Eq.~(\ref{eq:11}), or a similar functional dependence $\rho(a)$ that 
characterizes the perfect fluid (for example a non-linear equation of 
state). Therefore, when we refer to ``the Friedmann equation'' 
we mean Eq.~(\ref{eq:11}) {\em plus} this extra ingredient, which makes 
the analogy complete. All the equations for physical and geometrical  
problems considered in the following have  a form that already includes  
some familiar dependence $\rho(a)$ of the fluid energy density on the 
scale factor and makes it suitable for a complete analogy.

In the following we review the most celebrated textbook problems of 
variational calculus in one dimension, building cosmological analogies 
where possible.

\section{Geodesics of the Euclidean plane and a not-so-trivial analogue}
\setcounter{equation}{0}
\label{sec:2}

A simple variational problem consists of finding the geodesic curves 
extremizing the length between two fixed points in the Euclidean plane. 
The infinitesimal arc 
length along a curve $y(x)$ in this plane is 
$dl=\sqrt{dx^2+dy^2}=\sqrt{1+y'^2} \, dx$, 
where $y' \equiv dy/dx$. The finite length betwen two fixed points along 
this curve is the functional of the curve
\be 
J\left[ y(x) \right]=\int_1^2 dl=\int_{x_1}^{x_2} dx \, \sqrt{1+y'^2} 
\equiv 
\int_{x_1}^{x_2} dx \, L\left( y'(x) \right) \,.
\ee
Since the Lagrangian $L$ does not depend  on $y$, the canonically 
conjugated momentum $\partial L/\partial y'$ is conserved, 
\be
\frac{y'}{\sqrt{1+y'^2}}= C
\ee
or
\be
y'=\frac{C^2}{1-C^2} 
\ee
which integrates trivially to $y(x)= \frac{C^2}{1-C^2} \, x + D$, 
giving straight lines as the 
geodesics of the Euclidean plane. This equation gives also 
\be
\left( \frac{y'}{y} \right)^2 = \frac{C_1^2}{y^2}  
\ee
(where $C_1^2=C^2/(1-C^2)$) and the analogous Friedmann 
equation~(\ref{eq:11}) is
\be
H^2 = \frac{C_1^2}{a^2}  \,.
\ee
It describes an empty 
cosmos with hyperbolic ($K=-1$) spatial 
sections, 
known as the Milne universe. This solution of the Einstein-Friedmann 
equations is nothing but Minkowski spacetime in disguise because of a 
hyperbolic foliation ({\em i.e.}, in accelerated coordinates). In these 
coordinates, flat spacetime is sliced with negatively curved spatial 
sections, all the components of the Riemann tensor vanish identically, but 
the intrinsic curvature of the spatial 3-sections does not  
\cite{Mukhanov}.

The function $H(a)$ is given in Fig.~\ref{fig-H(a)}.

\section{The catenary problem}
\setcounter{equation}{0}
\label{sec:3}

Consider a heavy string 
hanging in a vertical $\left( x, y \right)$ plane and 
described by the profile $y(x)$. The linear density is 
$\mu=dm/dl$, where $dl=\sqrt{dx^2+dy^2}=\sqrt{ 1+(y')^2}\, 
dx $ is the elementary arc length along the string.  The 
gravitational potential energy of an element of string of 
length $dl$ located at horizontal position $x$ is $dE_g= 
\mu g y(x) ds $. The total gravitational potential energy of 
a string suspended by two points of horizontal coordinates 
$x_1$ and $x_2$ is the functional of the curve $y(x)$
\be
E_g \left[ y(x) \right] = \mu g \int_{x_1}^{x_2} dx \, y 
\sqrt{1+(y')^2} \equiv \int_{x_1}^{x_2} L \, dx 
\,.
\ee
The Lagrangian $L\left( y(x), y'(x) \right) $ does not 
depend explicitly on the coordinate $x$ and, therefore, the 
corresponding Hamiltonian is conserved:
\be
{\cal H} =\frac{\partial L}{\partial y'} \, y'-L=c_1 
\,,\label{eq:azzolina}
\ee 
where $c_1$ is a constant. Equation~(\ref{eq:azzolina}) (known as the 
Beltrami 
identity) simplifies to 
\be \label{appendix-questa}
\frac{-y}{\sqrt{ 1+(y')^2}}=c_1 \,,
\ee
which has catenaries as solutions \cite{Goldstein}. 

Although not necessary, the variational principle is 
sometimes imposed subject to the constraint that the  
string length between $x_1$ and $x_2$ is fixed, 
which changes the variational integral to 
\be
J \left[ y(x) \right] = \int_{x_1}^{x_2} dx \, \left( y+\lambda \right) 
\sqrt{1+(y')^2} \,,
\ee
where $\lambda $ is a Lagrange multiplier. A shift $y\rightarrow 
\bar{y}\equiv y+\lambda$ reduces this problem to the previous one.

The cosmological analogue of Eq.~(\ref{appendix-questa}) 
corresponds to a well known situation in cosmic physics. This equation is 
recast as 
\be
\left( \frac{ y'}{y} \right)^2 = \frac{1}{C_1^2}-\frac{1}{y^2} \,,
\ee
the analogue of which for the scale factor $a(t)$ reads
\be
H^2 = \frac{\Lambda}{3}-\frac{1}{a^2} \,,
\ee
where $\Lambda= 3/C_1^2 >0$ is the cosmological constant. We continue the 
discussion in the next problem, which leads to the same cosmic analogue.

\section{Minimal surface of revolution and its analogue}
\setcounter{equation}{0}
\label{sec:4}

Another classic problem of variational calculus is that of the  minimal 
surface of revolution. Let $y(x)$ join two points in the vertical $\left( 
x, y \right)$ plane and rotate this curve about the vertical (or $y$-) 
axis. 
The problem of finding the curve that achieves the surface of minimal area 
is solved by extremizing the area integral \cite{Goldstein}
\be
J =\int_{x_1}^{x_2} dx \, x \sqrt{1+\left( 
\frac{dy}{dx} \right)^2}  \,.
\ee
A practical application is given by soap bubbles between wire frames. 
Since the energy of a soap bubble is proportional to its area, nature 
tends to minimize it and to achieve the minimal surface of revolution 
when soapy water is placed on a wire frame obtained by rotating about the 
$y$-axis a wire shaped as the graph of $y(x)$.

\subsection{Dependent variable $x=x(y)$}

It is now convenient to take $x(y)$ as the dependent 
variable instead of $y(x)$ and to rewrite the integral as
\be
J\left[ x(y) \right]=\int_{y_1}^{y_2} dy \, x \sqrt{1+\left( \frac{dx}{dy} 
\right)^2}  \equiv 
\int_{y_1}^{y_2} dy \, L\left( x(y), \frac{dx}{dy} \right) \,,
\ee
where $y_{1,2} \equiv y (x_{1,2})$. Since $\partial L/\partial y =0$ the 
corresponding Hamiltonian is conserved,
\be
{\cal H}= \frac{ \partial L}{\partial x'} \, x'-L= \frac{x x'^2}{ 
\sqrt{1+x'^2}} -x \sqrt{1+x'^2} = C \,,
\ee
where $C$ is a  constant and now $x' \equiv dx/dy$. Manipulation of this 
equation yields
\be
x'^2=\frac{x^2}{C^2} -1 \,,\label{eq:minimalsurface}
\ee
which is solved by separation of variables, leading to
\be
\int \frac{dx}{\sqrt{x^2-C^2}} = \ln \left(\sqrt{x^2-C^2}+x \right)  = 
\frac{y-y_0}{C} 
\ee
where $y_0$ is an integration constant. By exponentiating both sides, a 
little algebra gives easily the solution
\be
x(y) = \frac{1}{2} \left( \mbox{e}^{\frac{y-y_0}{C} } + C^2 
\mbox{e}^{-\, \frac{y-y_0}{C} } \right) \,.
\ee
The initial conditions $C=\pm 1$ yield the catenary curve
\be
x(y)= \cosh \left(y-y_0 \right)
\ee
and $y_0=\cosh^{-1} x_0$ where $x_0 \equiv x(0)$.

The cosmic analogue is obtained by rewriting 
Eq.~(\ref{eq:minimalsurface}) as
\be
\left( \frac{x'}{x} \right)^2 = \frac{1}{C^2} -\frac{1}{x^2} \,.
\ee
The analogous Friedmann equation~(\ref{eq:11}) is 
\be
H^2 = \frac{\Lambda}{3} -\frac{1}{a^2} \,,
\ee
the right hand side of which contains a contribution from the positive 
cosmological constant $\Lambda=3/C^2$ and a contribution from the 
curvature term $-K/a^2$ with  $K=+1$. The analogous universe is well known 
as 
the positively curved universe with no matter and a pure (positive) 
cosmological constant; it evolves with scale factor
\be
a(t) = \sqrt{ \frac{3}{\Lambda} } \, \cosh \left[ \sqrt{\frac{\Lambda}{3}} 
\, \left(t-t_0 \right) \right] \,.
\ee
This catenary history of the universe includes a contracting 
phase (eternal in the past) for 
$t<t_0$, a bounce (made possible by the fact that the cosmological 
constant 
violates the strong energy condition and avoids the singularity) at 
$t=t_0$, followed by external expansion for $t>t_0$. The universe  
asymptotes to the de Sitter space with scale factor 
\be
a_{dS}(t) =  \sqrt{ \frac{3}{4\Lambda} } \, \exp \left[ 
\sqrt{\frac{\Lambda}{3}} 
\, \left(t-t_0 \right) \right] 
\ee
as $t\rightarrow +\infty$.

\subsection{Dependent variable $y=y(x)$}

In this case the Lagrangian is
\be 
L\left( x, y'(x) \right)= x\sqrt{1+y'^2} \,;
\ee
since $\partial L/\partial y=0$ the canonically conjugated momentum is 
conserved,
\be
\frac{\partial L}{\partial y'}= \frac{xy'}{\sqrt{ 1+y'^2}} = C\,.
\ee
This equation leads to 
\be
y'=\frac{C}{\sqrt{x^2-C^2} } \,,
\ee
which integrates to $y(x)= C\ln \left( \sqrt{x^2-C^2}+x \right) + 
\mbox{const}.$ 
and manipulations lead again to $ x(y)=\cosh\left( y-y_0\right)$ if $C=\pm 
1$. There is no cosmic analogue with this choice of variable because the 
putative analogue of the Friedmann equation
\be
\frac{y'^2}{y^2} = \frac{C^2}{y^2 \left(x^2-C^2\right)} 
\ee
contains $x$ explicitly, contrary to the Friedmann 
equation~(\ref{eq:11}) that does not contain $t$ explicitly. 

The function $H(a)$ is given in Fig.~\ref{fig-H(a)}, while the 
scale factor $a(t)$ is reported in Fig.~\ref{a-fig1}.

\section{Cosmic analogue of the brachistochrone problem}
\setcounter{equation}{0}
\label{sec:5}

The brachistocrone is the curve in the vertical plane that connects two 
given points (not on the same vertical) and such that a particle sliding 
on it without friction arrives at the bottom in the minimum time. The 
classic problem of finding this curve was posed to the elite mathematical 
community by Johann Bernoulli in 1696 \cite{Bernoulli}.

The speed $v$ of a particle falling from rest ($v_0=0$) from the height 
$y$ is given by energy 
conservation,
\be
\frac{mv^2}{2} = mgy 
\ee
(where  $g$ is the constant acceleration of gravity), from which one 
obtains the well known result $v=\sqrt{2gy}$. The 
descent time from point $1$ to point $2$ along the curve $y(x)$ with arc 
length $dl=\sqrt{dx^2+dy^2}$ is 
\be
J\left[ y(x) \right] =\int_1^2 dt= \int_1^2 \frac{dl}{v}= \int_{x_1}^{x_2} 
dx \, 
\sqrt{ \frac{1+y'^2}{2gy} } \equiv \int_{x_1}^{x_2} dx \, L\left( y(x), 
y'(x) \right) \,,
\ee
where $y(x)$ are trajectories 
from $1$ to $2$ in the vertical 
$\left(x,y\right)$ plane, and a 
prime denotes differentiation with respect to $x$. Clearly, it must be 
$y>0$. The descent time is extremized, $\delta J=0$, if $L$ satisfies the 
Euler-Lagrange equation. 

Since $\partial L/\partial x=0$ the corresponding Hamiltonian is 
conserved,
\be
{\cal H}=\frac{\partial L}{ \partial y'} \, y' -L=C_0 \,,
\ee
where $C_0$ is a constant. This Beltrami identity can be written as 
\be
\sqrt{y\left(1+y'^2\right)} = C_1 
\ee
(with $C_1=-1/(C_0\sqrt{2g}) >0$), from which one obtains
\be
y'^2=\frac{C_1^2}{y}- 1 \,, \label{eq:brachi}
\ee
which has a cycloid (the prototypical roulette) as its well known 
solution \cite{Goldstein}. One can 
divide both sides 
of Eq.~(\ref{eq:brachi}) by $y^2$ to obtain
\be
\left( \frac{y'}{y} \right)^2 = \frac{C_1^2}{y^3} - \frac{1}{y^2} \,,
\ee
which is analogous to the Friedmann equation
\be
H^2 = \frac{8\pi G \rho_0}{3a^3} - \frac{1}{a^2} 
\ee
describing a closed ($K=+1$) FLRW universe filled with a fluid with 
equation of state parameter $w=0$, {\em i.e.}, a dust with energy density 
scaling 
as 
$\rho=\rho_0 /a^3$. It is significant that the energy density comes 
out positive, which could spoil the analogy if it wasn't true and is not 
to be 
taken for granted when building analogies. This is a very simple 
matter-dominated universe and 
a classic textbook case. The solution is expressed in parametric form by
\begin{eqnarray}
a(\eta) & = & \frac{C_2}{2}\left(1-\cos \eta \right) \,, \\
&&\nonumber\\
t(\eta) & = &  \frac{C_2}{2}\left(\eta -\sin \eta \right) \,,
\end{eqnarray}
where the parameter $\eta$ is the conformal time, the initial condition 
$a(t=0)=0$ has been imposed, and $C_2=8\pi G \rho_0$.

On the mechanical side of the analogy, the parameter $\eta$ (the analog of 
conformal time)  is defined by 
$dx=yd\eta$, which means that small increments of $\eta$ are small 
increments of the coordinate $x$ measured in units of the height of the 
point lying on the brachistrochrone curve.

Equation~(\ref{eq:brachi}) shows that $y'\rightarrow \infty$ as 
$y\rightarrow 0$: the curve must start vertical, with a cusp, at its 
highest point. On the cosmology side of the analogy, this peculiarity 
corresponds to the fact that a universe filled with dust and positively 
curved necessarily begins at a Big Bang singularity; this is a special 
case of the general Hawking-Penrose singularity theorems satisfied by 
matter obeying the null energy condition (in this case, dust satisfies 
also the strong, weak, and dominant energy conditions \cite{Wald}).

 The function $H(a)$ is given in Fig.~\ref{fig-H(a)}, while the 
scale factor $a(t)$ is reported in Fig.~\ref{a-fig2}.

\section{Geodesics of the Poincar\'e half-plane}
\setcounter{equation}{0}
\label{sec:6}

The Poincar\'e half-plane is the upper part of the $\left(x, y \right)$ 
plane with $y>0$ and the metric given by the line element
\be
ds^2 =\frac{1}{y^2} \left( dx^2+dy^2\right)
\ee
conformal to the Euclidean metric. The arc length along a curve $y(x) $ is 
$dl=\sqrt{dx^2+dy^2}/y $ and geodesics are found by extremizing the 
finite length 
\be
l\left[ y(x)\right] =\int_{x_1}^{x_2} dx \, \frac{\sqrt{1+y'^2}}{y} \equiv 
\int_{x_1}^{x_2}  dx \, L\left( y, y'\right) 
\ee
between two given points.  Since $\partial L/\partial x=0$ the 
corresponding Hamiltonian is conserved,
\be
\frac{\partial L}{\partial y'} \, y'-L= C <0 \,,
\ee
which yields
\be
\frac{1}{Cy}=-\sqrt{1+y'^2} \,.\label{zx}
\ee
The solutions for $C=-1$ are half-circles perpendicular to the $x$-axis.  

The cosmic analogue obtained from Eq.~(\ref{zx})
\be
H^2= \frac{1}{C^2a^4} -\frac{1}{a^2} 
\ee
describes a positively curved ($K=+1$) universe filled with a  
radiation fluid with equation of state $P=\rho/3$ and $ 
\rho=\rho_0/a^4$, $\rho_0 = \frac{3}{8\pi G C^2}$. This is another 
classic textbook example with scale factor
\be
a(t)=\sqrt{C'} \sqrt{ 1-\left( 1-\frac{t}{C'} \right)^2} \,,
\ee
where $C'=1/C^2$. This analogy was already discussed in Rindler's textbook 
\cite{Rindler} and in Ref.~\cite{CriadoAlamo}.

The function $H(a)$ is given in Fig.~\ref{fig-H(a)}, while the 
scale factor $a(t)$ is reported in Fig.~\ref{a-fig1}.

\section{The gravity tunnel}
\setcounter{equation}{0}
\label{sec:7}

A popular problem in introductory physics courses consists of analyzing 
the motion of a particle through a tunnel dug out along a diameter of the 
Earth, which is modelled as a uniform sphere \cite{Routh, Cooper, Kirmser, 
Venezian, Mallett, Laslett} (see \cite{Klotz} for a non-uniform sphere). A 
homogeneous 
sphere of mass $M$ and radius $R$ has density 
$\rho_s=\frac{3M}{4\pi R^3}$ and the spherically symmetric Newtonian 
potential $\Phi(r)$ satisfies the Poisson equation
\be
\frac{1}{r^2} \, \frac{d}{dr} \left( r^2 \, \frac{d\Phi}{dr} \right)= 
4\pi G \rho_s 
\ee
inside the sphere. This readily integrates to
\be
\Phi(r) =\frac{GM}{2R^3} \, r^2 -\frac{C_1}{r} +C_2 \,,
\ee
where $C_{1,2}$ are integration constants. Regularity at the centre 
requires $C_1=0$ and matching the exterior potential  $\Phi_{out}=-GM/r$ 
at the surface $r=R$ yields $C_2=-\, 
\frac{3GM}{2R}$. Therefore, the potential inside the homogeneous sphere
\be
\Phi(r)= \frac{GM}{2R^3} \, r^2 -\frac{3GM}{2R} 
\ee
is that of  a shifted harmonic oscillator and the particle oscillates 
up and down the tunnel.

The Lagrangian for a particle of mass $m$ and position $r(t)$ moving 
without friction through this gravity tunnel is 
\be 
L\left( r, \dot{r} \right) = \frac{m\dot{r}^2}{2} -\frac{GM}{2R^3} \, r^2 
+\frac{3GM}{2R} \,; 
\ee 
since $\partial L/\partial t=0$ the corresponding 
Hamiltonian (which coincides with the particle energy) is conserved, 
\be 
\frac{\partial L}{\partial \dot{r}} 
\,\dot{r} - L =E \,, 
\ee 
expressing the energy integral  
\be 
\frac{m \dot{r}^2}{2} +m \left( \frac{GM}{2R^3} 
\, r^2 -\frac{3GM}{2R} \right)=E \,. 
\ee 
This Beltrami identity can be rearranged as 
\be 
\left( \frac{\dot{r}}{r} \right)^2 = -\frac{GM}{R^3} -\frac{K}{r^2} 
\,, 
\ee 
where $K=-\, \frac{2}{m} \left( E+\frac{3GM}{R} \right)$. The 
analogous Friedmann equation~(\ref{eq:11}) for $a(t) \leftrightarrow r(t)$ 
is 
\be 
H^2= \frac{\Lambda}{3} -\frac{K}{a^2} \label{eq:AdS} 
\ee 
and it 
exhibits a negative cosmological constant $\Lambda=-3GM/R^3$. There is no 
matter fluid in this analogous universe.  Since $H^2\geq 0$, only a 
negative negative curvature index 
$K<0$ is possible to compensate for the negative $\Lambda$. Apart from 
the dimensionality, this universe 
is the anti-de Sitter spacetime of the AdS/CFT correspondence and the 
holographic 
principle \cite{AdSCFT}. The solution of Eq.~(\ref{eq:AdS}) is the scale 
factor 
\be 
a(t) = \sqrt{ \frac{3}{|\Lambda|} } \, \sin\left( \sqrt{ 
\frac{|\Lambda|}{3} } \, t \right) \,. 
\ee

 The function $H(a)$ is given in Fig.~\ref{fig-H(a)}, while the 
scale factor $a(t)$ is reported in Fig.~\ref{a-fig1}.

\section{The terrestrial brachistochrone}
\setcounter{equation}{0}
\label{sec:8}

In the terrestrial brachistochrone 
problem \cite{Cooper, Venezian}, tunnels of various curved shape  
described by functions $r(\theta)$ in polar coordinates, are dug out in 
the uniform Earth of radius $R$, and one looks for the shape that 
minimizes the transit time between two given points. This problem has seen 
much renewed attention recently, mostly in the pedagogical literature 
\cite{new1, new2, new3, new4, new5, new6, new7, new8, new9, new10, new11, 
new12, new13, new14} but not only \cite{app1, app2, app3}.

The corresponding Lagrangian is \cite{Venezian} 
\be L\left( r, r'\right)=\sqrt{ \frac{ r^2+r'^2}{R^2-r^2}} \,,
\ee
where $r' \equiv dr/d\theta$. Since the Lagrangian does not depend 
explicitly on $\theta$ the corresponding Hamiltonian is conserved, 
\be
{\cal H}=\frac{\partial L}{\partial r'} \, r'-L = C \,,
\ee
leading to 
\be
\frac{r^2}{ \sqrt{R^2-r^2} \, \sqrt{r^2+r'^2} }= -C \,.
\ee
Further manipulation gives
\be
\frac{r'^2}{r^2} = -1 +\frac{r^2}{C^2 \left( R^2-r^2\right)} \,.
\ee
The minimum radius is attained where the curve $r(\theta)$ flattens, 
$r'=0$, yielding
\be
C^2=\frac{r_\text{min}^2}{R^2-r_\text{min}^2 } \,,
\ee
then one can write
\be
\frac{r^2 +r'^2}{r^2} = \frac{r^2}{r_\text{min}^2} \, 
\frac{(R^2-r_\text{min}^2)}{R^2-r^2} \,.
\ee
The analogous Friedmann equation 
\be
H^2= -1 +\frac{a^2}{C^2 \left(a_0^2-a^2\right)} \label{TBFriedmann}
\ee
is more interesting than in the situations discussed previously. The 
universe contains a negative cosmological constant $\Lambda=-3$ and   
a perfect fluid with energy 
density 
\be
\rho=\rho_0 \, \frac{a^2}{a_0^2 - a^2} \,, \;\;\;\;\;\;\;\; \frac{8\pi 
G}{3} \,  \rho_0 = \frac{1}{C^2} \,.
\ee
By imposing the covariant conservation equation 
$\dot{\rho}+3H(P+\rho)=0$, one finds the equation of state of 
the cosmic fluid. First, we obtain
\be
P=-\rho -\frac{2\rho_0 a_0^2 \, a^2}{ 3\left( a_0^2-a^2 \right)^2 } \,;
\ee
then, using $ \frac{a^2}{(a_0^2-a^2)^2}= \left( 
\frac{\rho}{\rho_0}\right)^2 
\frac{1}{a^2}$ and $a^2 = 
\frac{\rho}{\rho_0} \, \frac{a_0^2}{(1+\rho/\rho_0)}$, one derives the 
quadratic barotropic  equation of state 
\be
P(\rho)=-\frac{5}{3}\, \rho -\frac{2}{3} \frac{\rho^2}{\rho_0} \,.
\ee
Since $P<-\rho$ this equation describes a phantom fluid \cite{Caldwell}. 
Equations of state of the cosmic fluid with the form $P=\sum_{k=1}^m c_k 
\rho_{(k)}^k$ have been studied recently in cosmology \cite{Gibbons0, 
roulettes, Szydlowski} as forms of dark energy. In particular, quadratic 
equations of state are 
the subject of several works \cite{quadratic, quadratic2, quadratic3, 
quadratic4, quadratic5, quadratic6}. These exotic equations of state 
produce peculiar types of spacetime singularities. While traditional 
cosmological solutions for linear barotropic equations of state contain 
Big Bang, Big Crunch, and Big Rip type singularities, the discovery of the 
acceleration of the universe in 1998 prompted the consideration of many 
more exotic non-linear equations of state for the dark energy fluid, which 
must be postulated in order to explain the cosmic acceleration within 
general relativity. This broadening of the picture results in a much wider 
spectrum of possible singularities \cite{BarrowSFS, BarrowGallowayTipler, 
Sahni, Bamba, Dabrowski2, Dabrowski, Fernandez, Szydlowski, Mariam, 
Diego}.

The Ricci scalar 
\be
{\cal R}= -\rho_{total}+3P_{total} = \frac{1}{2\pi G} -6\rho -2\, 
\frac{\rho^2}{\rho_0} 
\ee 
diverges as $a\rightarrow a_0$, signalling a spacetime singularity where 
the scale factor stays finite but $H, \rho$, and $P$ diverge. In 
the mechanical side of the analogy, $a_0$ corresponds to $R$, the boundary 
of the physical problem. On the cosmology side of the 
analogy, $a_0$ is a 
barrier that cannot be crossed dynamically:   $a(t)$ always remains 
smaller than $a_0$, but it approaches it, as will soon be clear. The 
graph 
of $a(t)$ has a cusp where $a$ reaches 
$a_0$, as is evident from the Friedmann equation~(\ref{TBFriedmann}). In 
the analogy, this 
feature corresponds to the fact that the terrestrial brachistochrone 
starts at the surface ($r=R$, corresponding to $a=a_0$) perpendicular to 
it with infinite derivative $r'\equiv dr/d\theta$ \cite{Venezian, Klotz}.

The fact that there is a minimum value of $a(t)$ also follows immediately 
from Eq.~(\ref{TBFriedmann}). Since it must be $H^2 \geq 0$, the region 
$a< a_\text{min}$ is forbidden to the dynamics, where
\be
a_\text{min}^2= \frac{C^2a_0^2}{1+C^2} \,,
\ee
is the value of $a$ corresponding to $H=0$; therefore the dynamics is 
restricted to the strip of the $\left( t, a \right)$ plane  
\be
a_\text{min}\leq a(t)< a_0 \,.
\ee
We also have  
\be
C^2=\frac{3}{8\pi G \rho_0} = \frac{ a_\text{min}^2}{a_0^2-a_\text{min}^2} 
\,.\label{Csquared}
\ee
This minimum value of $a$ is the analogue of the minimum radius 
$r_\text{min}$ that can 
be reached, but not passed, by a particle in motion on the terrestrial  
brachistochrone (this curve  does not pass through the centre of the 
Earth) \cite{Venezian, Klotz}. 

The static universe with $a(t) \equiv a_\text{min}$ is an exact solution 
of the Einstein-Friedmann equations, corresponding to a balance between 
$\Lambda<0$ (which is attractive) and $(\rho + 3P) <0$, which is repulsive 
in the acceleration equation~(\ref{eq:12}), where these terms appear with 
opposite sign and compete. Since this equilibrium is fine-tuned, one 
expects the static solution to be unstable. Indeed, a linear perturbation 
analysis reveals an exponentially growing mode (see the Appendix). This 
static solution is the only fixed point of the dynamics and is unstable 
and a repellor.

Further, the acceleration equation~(\ref{eq:12}) gives $\ddot{a}>0 $ for 
all values of 
the scale factor $a_\text{min}< a(t) <a_0$; the concavity of $a(t)$ faces 
upward, its derivative $\dot{a}$ increases, and is larger and larger the 
closer the orbit of the solution gets 
to the singularity $a_0$.  The picture of the dynamics that emerges from 
these considerations is the following: solutions starting anywhere in the 
strip $a_\text{min} <a <a_0$ move toward the singularity faster and faster, 
with increasing ``speed'' $\dot{a}$, until they hit it and the universe 
ends in a finite proper time, with this finite value $a_0$ of the scale 
factor.

The analogy that we have created provides an explicit example of a 
universe dominated by a  phantom fluid with quadratic equation of state. 
This analogy is useful since the exact solution of the terrestrial 
brachistochrone is known \cite{Venezian, Klotz}. It is  a hypocycloid, the 
curve generated by a circle of radius $\left( R-r_\text{min}\right)/2$ 
rolling without sliding at constant speed inside the larger circle of 
radius $R$ (clearly, this curve is another roulette).  
It can be given in 
parametric form as 
\begin{eqnarray}
r^2(t) &=& \frac{R^2+r_\text{min}^2}{2} 
-\frac{\left(R^2-r_\text{min}^2\right)}{2} \, \cos \left( 
2\omega \, t\right) \,,\\
 &&\nonumber\\
\theta(t) &=& \tan^{-1}\left[ \frac{R}{r_\text{min}} \, \tan (\omega \, t) 
\right] -\frac{r_\text{min}}{R} \, \omega \, t \,,
\end{eqnarray}
where time is the parameter, $\omega=2\pi/T$, and 
\be
T=\pi \sqrt{ \frac{R^2-r_\text{min}^2}{Rg} }
\ee
(with $g$ the acceleration of gravity at the surface) is the travel time 
between two points on the surface of the Earth, which is the minimum 
travel time among all tunnel configurations between these 
two points \cite{Venezian, Klotz}. The 
solution can  be written also as \cite{Venezian, Klotz} 
\begin{eqnarray}
\theta(r) &=& \tan^{-1} \left( \frac{R}{r_\text{min}} \, \sqrt{ 
\frac{r^2-r_\text{min}^2}{R^2-r^2}} \right) 
-\frac{r_\text{min}}{R} \,  \tan^{-1} \left( \sqrt{ \frac{ 
r^2-r_\text{min}^2 }{R^2 
-r^2}} \, \right) \,,\\
&&\nonumber\\
t(r) &=& \frac{ \sqrt{R^2-r_\text{min}^2} }{2\sqrt{ Rg} } \, \cos^{-1} \left( 
\frac{R^2+r_\text{min}^2-2r^2}{R^2-r_\text{min}^2} \right) \,.
\end{eqnarray}
The exact solution can immediately be transposed into the solution for the 
analogous universe
\begin{eqnarray}
a^2(s) &=& \frac{a_0^2+a_\text{min}^2}{2} -\frac{\left( 
a_0^2-a_\text{min}^2 \right)}{2} \cos \left( 2\omega \, s \right) 
\,,\label{para1}\\
 &&\nonumber\\
t(s) &=& \tan^{-1}\left[ \frac{a_0}{a_\text{min}} \, \tan (\omega \, s) \right] 
-\frac{a_\text{min}}{a_0} \, \omega \, s \,,\label{para2}
\end{eqnarray}
or
\begin{eqnarray}
t(a) &=& \tan^{-1} \left( \frac{a_0}{a_\text{min}} \, \sqrt{ 
\frac{a^2-a_\text{min}^2}{a_0^2-a^2}} \right) 
-\frac{a_\text{min}}{a_0} \, \tan^{-1} \left( \sqrt{ \frac{ a^2-a_\text{min}^2 
}{a_0^2 -a^2}} \, \right) \,,\\
&&\nonumber\\
s(a) &=& \frac{ \sqrt{a_0^2-a_\text{min}^2}}{2\sqrt{a_0 g}} \, \cos^{-1} \left( 
\frac{a_0^2+a_\text{min}^2-2a^2}{a_0^2-a_\text{min}^2} \right) \,.
\end{eqnarray}
The singularity $a=a_0$ is approached, and the universe ends its 
existence,  in a finite time. In fact,  the limit $a\rightarrow a_0$ in 
the  parametric solution (\ref{para1}), (\ref{para2}) yields the parameter 
value $s_0=\pi \, \sqrt{ 
\frac{a_0^2-a_\text{min}^2}{a_0 g}}$ and the finite time
\be
t_0= \frac{\pi}{2} \left( 1-\frac{a_\text{min}}{a_0} \right) \,.
\ee
By equating $t_0$ with $T/2$, we obtain
\be
g=a_0 \left( \frac{a_0+a_\text{min}}{a_0-a_\text{min}} \right) 
\ee
in the cosmic analogue. 

\begin{figure}
\centering
\includegraphics[width=10 cm]{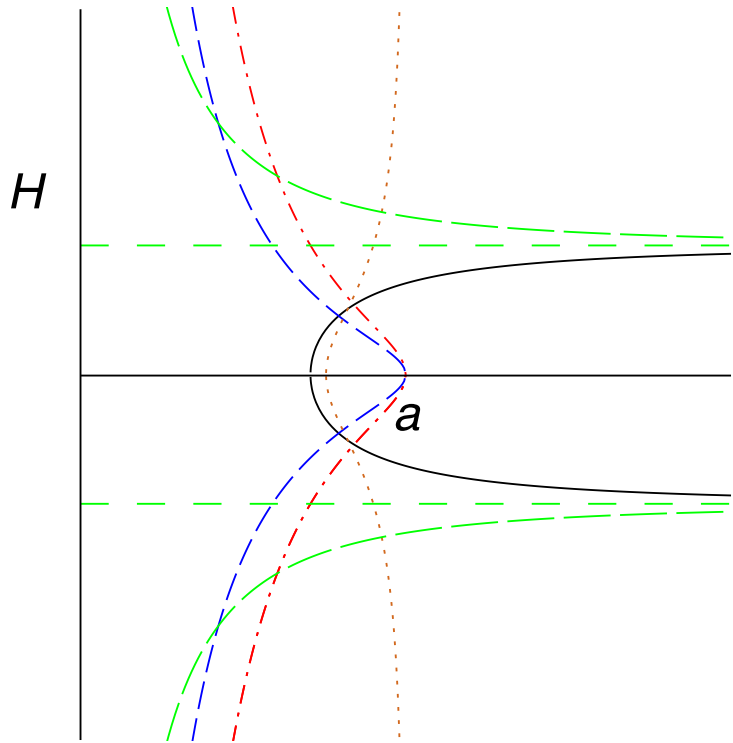}
\caption{The functions $H(a)$ for the problems discussed. The  
curves correspond, respectively, to the catenary/minimal surface of 
revolution problems 
(black, solid); the brachistochrone problem (blue, dashed); the 
problem of the Poincar\'e 
half plane geodesics (red, dash-dotted); the gravity tunnel problem for 
$K=-1, 
0$ (green, long-dashed) or $K=1$ (black, solid); and the terrestrial 
brachistochrone problem (chocolate, dotted).\label{fig-H(a)}} 
\end{figure}

\begin{figure}
\centering
\includegraphics[width=10 cm]{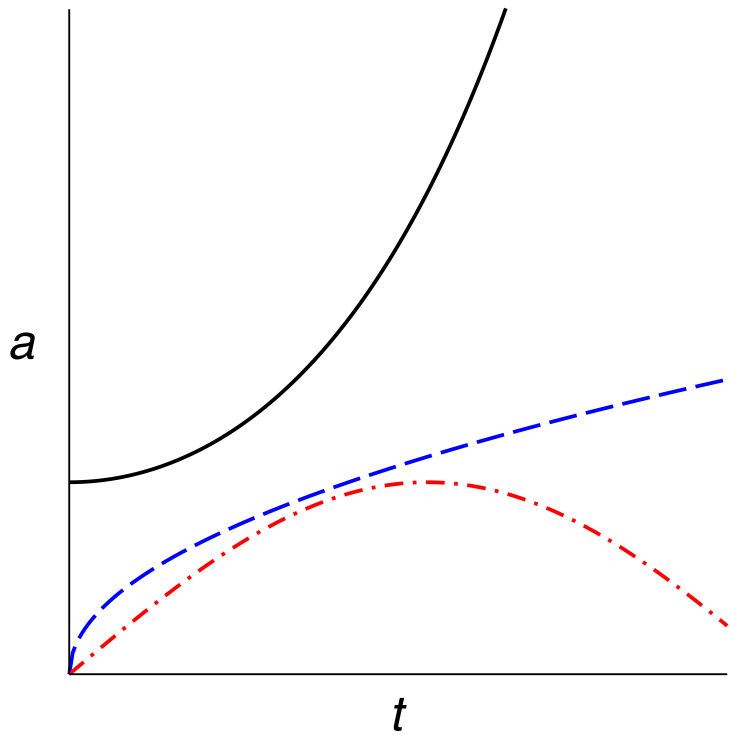}
\caption{The scale factor $a(t)$ for the problems discussed. The 
curves correspond, respectively, to the catenary/minimal surface of 
revolution problems 
(black, solid); the Poincar\'e 
half plane geodesics problem (blue, dashed); and the gravity tunnel 
problem
 (red, dash-dotted). \label{a-fig1}} 
\end{figure}

\begin{figure}
\centering
\includegraphics[width=10 cm]{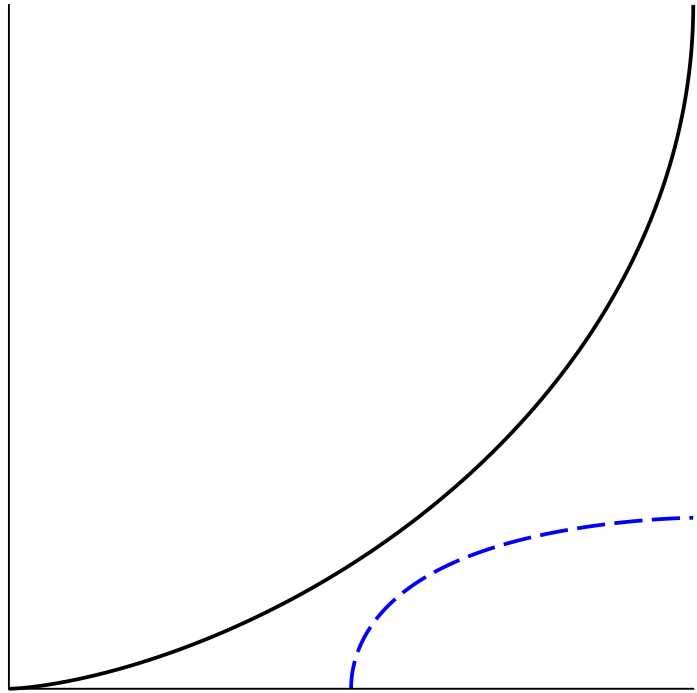}
\caption{The scale factor $a(t)$ for the remaining problems discussed. 
The curves correspond, respectively, to the brachistochrone problem  
(black, solid) and the terrestrial brachistochrone problem (blue, 
dashed).\label{a-fig2}} 
\end{figure}

 The function $H(a)$ is given in Fig.~\ref{fig-H(a)}, while the 
scale factor $a(t)$ is reported in Fig.~\ref{a-fig2}.

\section{Discussion and conclusions}
\setcounter{equation}{0}
\label{sec:9}

Several classic problems of variational calculus originating in 
one-dimensional non-relativistic particle mechanics have solutions that 
constitute analogues of spatially homogeneous and isotropic universes 
because they are ruled by an equation which is formally a Friedmann 
equation for a suitable perfect fluid. This property suggest that it is 
possible to build mechanical analogues of cosmological spacetimes.
If one begins with a universe with a given curvature index $K$ and 
equation of state $P=w(a) \rho$, the Friedmann equation~(\ref{eq:11}) 
looks 
like an energy conservation equation and immediately suggests the 
point-like Hamiltonian
\be 
{\cal H}= \frac{\dot{a}^2}{2} -\frac{4\pi G}{3} \, \rho a^2 
\ee
(which is constant and equal to $-K/2$) and the point-like Lagrangian
\be
L\left( a, \dot{a}\right)= \frac{\dot{a}^2}{2} +\frac{4\pi G}{3} \, \rho 
a^2 \,.
\ee
One can then compare this Lagrangian and Hamiltonian with that of 
an (otherwise unrelated)  
mechanical or geometrical system, as we have mostly done throughout this 
manuscript. Alternatively, one can compare directly a first order 
integral  for the system with the Friedmann equation~(\ref{eq:11}), or 
compare the 
second order 
equation of motion with the acceleration equation~(\ref{eq:12}). 

We make no
claim that the analogies proposed are full physical analogies. They may be 
limited to 
the mathematics, {\em i.e.}, formal analogies, as in the classic analogy 
between 
(forced and damped) mechanical oscillators and RLC electric circuits
reported in \cite{Goldstein}  and in most mechanics textbooks. There,
the analogy between the equations governing the two systems is
mathematical but, the equations being the same, they also have the same
oscillatory solutions.  This is a mathematical analogy and the two systems
are completely different from the physical point of view. However, if one
thinks about oscillations in nature in a general sense, the analogy has  
also a physical aspect. This
is a physical analogy only on an abstract level and we won't go as far as  
claiming it is a full physical analogy (which would bring us into 
semantics).

We have examined several situations: while most of them give rise to 
simple, albeit very important, FLRW solutions corresponding to simple 
fluids such as dust or radiation or a pure cosmological constant 
$\Lambda$, others correspond to phantom fluids \cite{Caldwell} with 
non-linear equations of state originating exotic singularities at finite 
time, where the scale factor remains finite while curvature invariants 
blow up. These types of singularities are studied and classified in recent 
literature \cite{BarrowSFS, BarrowGallowayTipler, Sahni, Bamba, 
Dabrowski2, Dabrowski, Fernandez, Szydlowski, Mariam, Diego}. The 
situations examined are summarized in Table~\ref{table:1}.

\small
\begin{table}
\centering
\begin{tabular}{ccc}
%\toprule
\textbf{Variational problem} & \textbf{Lagrangian} & \textbf{Cosmic analogue}\\
%\midrule
\hline\hline\\
Geodesics of $\mathbb{R}^2$	     & $L=\sqrt{1+y'^2}$	& $K<0, \rho=0$\\
	                     &               	        & Milne universe\\
                 		& 			&   \\
\hline\\
Catenary problem	     & 	$L=y\sqrt{1+y'^2}$	& $\Lambda>0, K>0,\rho=0$\\
                 		& 			&   \\
\hline\\
Minimal surface of revolution & $L=x\sqrt{1+x'^2}$	& same as above \\
                 		& 			&   \\
                 	      & $L=x\sqrt{1+y'^2}$	& no cosmic analogue  \\
                 		& 			&   \\
\hline\\
Brachistochrone             & 	$L=\sqrt{\frac{1+y'^2}{y}}$  & $K>0$, dust\\
	 	            &			         & cycloid\\
                 		& 			&   \\
\hline\\
Geodesics of Poincar\'e plane  & $L=\frac{\sqrt{1+y'^2}}{y}$ & $K>0$, radiation\\
                 		& 			&   \\
\hline\\
Terrestrial tunnel & $L=\dot{r}^2-\frac{GM}{R^3} \, r^2 +\frac{3GM}{R}$  & 
$\Lambda<0, K<0, \rho=0$\\
				&  		& anti-de Sitter space\\
                 		& 			&   \\
\hline\\
Terrestrial brachistochrone & $L=\sqrt{ \frac{r^2 +r'^2}{R^2 -r^2}}$ & $\Lambda<0, 
K=0, P=-\frac{5\rho}{3} - \frac{2\rho^2}{3\rho_0}$ \\ 
                 		& 			& hypocycloid\\
                 		& 			&   \\
\hline\hline
\end{tabular}
\caption{Cosmic analogues of variational problems}\label{table:1}
\end{table}
\normalsize

Some of the variational problems that we have examined are centuries old 
and are textbook material. It is not surprising that formal analogies with 
the Friedmann equation~(\ref{eq:11}) arise, since the latter takes the 
form of the first integral of motion corresponding to energy conservation 
for a particle in one-dimensional motion, and a variety of solutions are 
possible corresponding to the three possible curvatures and the wide range 
of perfect fluid equations of state (even limiting oneself to the 
barotropic equation $P=w\rho$ with $w=$~const.). Indeed, it is always 
possible to construct an analogy between a FLRW space filled by perfect 
fluids and the one-dimensional motion of a particle of unit mass and 
position $a(t)$ in a suitable potential \cite{UrenaLopez}, but the latter 
may be contrived and not physically meaningful.

Cosmic analogues have been reported for fluids \cite{fluid1,fluid2, 
fluid3, fluid4, fluid5, fluid6, fluid7, fluid8, fluid9, fluid10, fluid11, 
fluid12, fluid13}, Bose-Einstein condensates \cite{acosmo1, acosmo2, 
acosmo3, acosmo4, analoguegeneral2, BEC2, BEC3, BEC4, BEC5, BEC6, BEC7, 
CM2}, glacial valleys \cite{Gibbons0, FACETS}, optics \cite{optical1, 
optical2, optical3, optical4, optical5, Gibbons0}, capillary fluid flow 
\cite{capillary}, soap bubbles \cite{CriadoAlamo,bubble2}, equilibrium 
beach 
profiles \cite{beach}, the freezing of bodies of water \cite{freezing}, 
Omori's law for the aftershocks following a main earthquake shock 
\cite{myOmori}, {\em etc.} It is rather surprising, however, that formal 
analogies of classic textbook variational problems with FLRW cosmology 
which, as seen above, are associated with important solutions of the 
Friedmann equation~(\ref{eq:11}), are not reported in the literature (with 
the exception of the Poincar\'e half-space in \cite{Rindler, 
CriadoAlamo}).

From the previous analysis it is clear that one-dimensional variational 
problems described by a Lagrangian $L\left( y(x), y'(x) \right)$ not 
depending explicitly on the independent variable $x$ can have a cosmic 
analogue, with the Friedmann equation~(\ref{eq:11}) corresponding to the 
conservation of the Hamiltonian ${\cal H}=\frac{\partial L}{\partial y'}\, 
y'-L$. However, for the cosmic analogy to begin making sense, the 
corresponding perfect fluid must have a non-negative energy density 
$\rho$, which cannot be taken for granted and dooms many would-be 
analogies (for all the situations discussed in this work, it is $\rho \geq 
0$). Moreover, since all the solutions of the Friedmann equations are 
roulettes \cite{roulettes}, only variational problems that have roulette 
solutions can give rise to a cosmic analogy. This fact seems to restrict 
considerably the range of possible solutions to the problem of building a 
cosmic analogue for a given one-dimensional mechanical system, and makes 
the existing analogies more valuable. Indeed, in our search for cosmic 
analogues we have encountered cycloids, hypocycloids, and catenaries, 
which are classic examples of roulettes. 

In certain cases when $\partial L/\partial x \neq 0$, it is still possible 
to 
switch the dependent and independent variables from $y(x)$ to $x(y)$ in 
the action integral (and, correspondingly, change the integration 
variable), thus changing its form, and obtain a meaningful cosmic 
analogy.\footnote{Switching 
dependent and independent variables does not always lead to a variational 
principle, for example in the case of the brachistrochrone.} In general, 
when both descriptions admit cosmic analogues, switching dependent and 
independent variables leads to inequivalent cosmologies (this is not 
always the case, for example the problem of geodesics in the Euclidean 
plane leads to the same Milne universe in both descriptions, due to the 
simplicity of the Lagrangian $L=\sqrt{1+y'^2}$).

The problem of the terrestrial brachistochrone, which has seen renewed 
attention recently \cite{Klotz, new1, new2, new3, new4, 
new5, new6, new7, new8, new9, new10, new11, new12, new13, new14, app1, 
app2, app3}, provides an explicit example of a 
universe dominated by a phantom fluid with non-linear equation of state, 
which can be solved explicitly and exhibits a finite future singularity at 
a finite value of the scale factor, where the Hubble function, Ricci 
scalar, energy density, and pressure all diverge. Finite time 
singularities have been the subject of much literature in the past decade 
\cite{BarrowSFS, BarrowGallowayTipler, Sahni, Bamba, Dabrowski2, 
Dabrowski, Fernandez, Szydlowski, Mariam, Diego} hence, in this problem, 
the mechanical side of the analogy helps the cosmology side in the 
sense that the known exact solution for the terrestrial brachistochrone 
problem can be immediately translated into an analytical solution of the 
corresponding cosmology with complicated (non-linear) equation of state.

These cosmological analogies are sometimes useful (this was the case for 
the analogy between equilibrium beach profiles, which generated solutions 
unknown to the oceanography community \cite{beach}; other times they are 
not) and they stimulate the imagination, which in the end may turn out 
to be their most valuable feature.

The further search for one-dimensional roulette analogues of FLRW 
cosmologies and the study of their small fluctuations (possibly mimicking 
cosmological perturbation theory) will be pursued in the future.

\small
\section*{Acknowledgments} The author thanks Dr. Stefano Bellucci for the 
invitation to write this work, which is supported by the Natural Sciences 
and Engineering Research Council of Canada (grant No. 2016-03803) and by 
Bishop's University.
\normalsize

\appendix\newpage%\markboth{Appendix}{Appendix}
\renewcommand{\thesection}{\Alph{section}}
\numberwithin{equation}{section}
\section{Appendix: Instability of the static universe analogous to the 
terrestrial brachistochrone}
\setcounter{equation}{0}

%\appendix
%\section*{Instability of the static universe analogous to the 
%terrestrial brachistochrone}
%\setcounter{equation}{0}

Perturb the static universe solution so that $a(t)=a_\text{min}+\delta a (t)$ 
with $\delta a \geq 0$ since it can only be $a_\text{min} \leq a(t)< a_0$. The 
acceleration equation~(\ref{eq:12}) yields
\be
1-\frac{4\pi G}{3} \left( \rho_\text{min}+3P_\text{min}\right)=0 
\label{zerorder}
\ee
to zero order (where $\rho_\text{min}$ and $P_\text{min}$ are the energy density and 
pressure corresponding to $a=a_\text{min}$). In general, Eq.~(\ref{eq:12}) can 
be written as
\be
\ddot{a} =\frac{a^3\left( 2a_0^2-a^2\right)}{C^2 \left( 
a_0^2-a_\text{min}^2\right)^2} +a \,.
\ee
Expanding to first order in $\delta 
a/a_\text{min} \ll 1 $, using Eq.~(\ref{Csquared}), and keeping only the zero 
order and the linear terms, one obtains
\be
\delta \ddot{a} \simeq \frac{C^2}{a_\text{min}} \left\{ 2a_0^2 -a_\text{min}^2 +
\left[ \frac{2C^2}{a_\text{min} } \left( 2a_0^2 -a_\text{min}^2 +1 \right) 
+\frac{1}{a_\text{min}}  +1 \right]\delta a \right\} +a_\text{min} \,.
\ee
Using the zero order equation~(\ref{zerorder}), one obtains the 
equation satisfied by the perturbations $\delta a $ to first order 
\be
\delta \ddot{a} =  \Omega^2 \, \delta a \,,
\ee
where 
\be
\Omega^2 =1+\frac{C^2}{a_\text{min}^2} \left[ 
2C^2 \left( 1+2a_0^2 -a_\text{min}^2 \right)+1 \right]
\ee
is positive because $a_0>a_\text{min}$. Therefore, the solution to linear 
order 
is 
\be
\delta a (t) =A\, \mbox{e}^{\Omega \, t} + B\, \mbox{e}^{-\Omega \, t} 
\ee
with $\Omega \in \mathbb{R}$ and it contains  a mode that diverges as time 
progresses. 
The static solution $a=a_\text{min}$ is unstable.

%%%%%%%%%%%%%%%%%%%%%%%%%%%%%%%%%%%%%%%%%%%%%%%%%%%%%%%%%%%%%%%%%%%%%%%%
   
\end{document}